\documentstyle{article}
\begin{document}
\centerline{\Large \bf Geometrization of the Lax Pair Tensors}
\vspace*{0.37truein}
\centerline{D. B{\u a}leanu \footnote{E-mail:
dumitru@cankaya.edu.tr, Institute of Space Science, P.O. Box, MG-23, R
76900, Magurele-Bucharest, Romania }}
\baselineskip=12pt
\centerline{\footnotesize\it Department of Mathematics and Computer Sciences, 
Faculty of Arts and Sciences,}
\baselineskip=10pt
\centerline{\footnotesize\it {\c C}ankaya University, 06531 Ankara, Turkey}
\vspace*{10pt}
\centerline{ S. Ba{\c s}kal
\footnote{E-mail: baskal@newton.physics.metu.edu.tr}}
\baselineskip=12pt
\centerline{\footnotesize\it Physics Department, Middle East Technical
University, Ankara, 06531, Turkey}
\vspace*{0.225truein}
\begin{abstract}
\noindent
The tensorial form of the Lax pair equations are given
in a compact and geometrically transparent form in the presence
of Cartan's torsion tensor.  Three dimensional spacetimes admitting Lax
tensors are analyzed in detail.  Solutions to Lax tensor equations include
interesting examples as separable coordinates and the Toda lattice.
\end{abstract}
\vspace*{-0.5pt}
\section{Introduction}
\vspace*{-0.5pt}
\noindent
In a series of papers, Rosquist et 
all.\cite{ros97,ros98,rosq3,patru,cinci} 
introduced the Lax tensors and presented some models for which Lax 
tensors exist.  The Lax representation for a given dynamical system
is not unique~\cite{lax68}.  Consequently, the Lax tensor equations
\begin{equation}
D_{\delta}L_{\alpha\beta\gamma}\,+\,D_{\gamma}L_{\alpha\beta\delta}
\,=\,L_{\alpha\mu(\gamma} B^{\mu}\,_{|\beta|\delta)}
-B_{\alpha\mu(\gamma} L^{\mu}\,_{|\beta|\delta)}
\label{one}
\end{equation}
depend on an arbitrary third rank
object $B^{\alpha}\,_{\beta\gamma}$, whose specific form can be
found through a suitable geometrization of the system.

The solutions of the Lax tensors were investigated on the
"dual" manifold\cite{holt96} in two dimensions~\cite{bk99}.
The connection between Lax tensors and Killing-Yano tensors of order
three has also been well-established~\cite{ros97}.  Killing-Yano
tensors of order three were introduced long time ago by 
Bocnher~\cite{bochner} and Yano~\cite{yano52} as a natural 
generalization of a Killing vector.  Gibbons et all.~\cite{gib}
found that Killing-Yano tensors can be understood as an object
generating a "non-generic symmetry", i.e., a supersymmetry
appearing only in the specific space-time.

An alternative way to introduce Lax tensors is to consider the
Killing and Killing-Yano tensors of order three.
Since the Killing-Yano tensors of order three are the generators of the
"non-generic" suppersymmetry on a given manifold~\cite{visine},
investigation of the manifolds admitting Lax tensors becomes
interesting.

Untill now a Killing tensor could only be found by solving the Killing
or the Killing-Yano equations~\cite{carter,holten95} of order
two, or by calculating the Nijenhuis tensor~\cite{das}.  The existence of
the Lax tensors of order three open a new possibility for finding Killing
tensors.

The three-particle open Toda lattice was geometrized by a suitable
canonical transformation and it was found that the tensor
$B^{\alpha}\,_{\beta\gamma}$ is antisymmetric with respect to
its first two indices~\cite{ros98}.
It is also known that the geometric duality can be generalized to
spinning spaces, at an expense of introducing a torsion on the
manifold~\cite{holt96}.  All these results suggest that torsion can
play an important role in the description of Lax tensors.

There are interesting questions yet to be answered as to the Lax
equations, and as to the geometrical interpretations of its
constituent tensors.  Specifically, the properties of the manifold
admitting Lax tensors deserve further investigation.
Although, on a very general context answers to these questions are not
easy to provide, problems concerning Lax tensor equations seem to be
manageable when some symmetry properties are imposed to its constituent
tensors, in some particular dimensions, which is what we intend to present
here.

This letter is organized as follows:~ In Sec. 2 the Lax tensor equations
are rederived in the presence of torsion to provide a geometric meaning
to the arbitrary tensor $B^{\alpha}\,_{\beta\gamma}$.
In Sec. 3 the Lax tensor equations are analyzed in three dimensions
for various symmetry properties of this tensor.
The integrability conditions of these equations are discussed for
specific cases such as flat and curved space-times as well as
spacetimes having torsion.  We present the explicit expressions
for the Lax tensors for a number of interesting examples including
separable orthogonal systems and the three-particle open Toda lattice.
The last section is devoted to our comments and conclusions.

\section{The tensorial Lax pair equations in the \\ presence of torsion}
\noindent
In this section we rederive the tensorial Lax pair equations
by taking into account a suitable definition for the Dirac-Poisson
brackets when torsion is introduced on an n-dimensional manifold.
The manifold is endowed with a metric
\begin{equation}
ds^{2}\,=\,g_{\mu\nu}dq^{\mu}dq^{\nu}
\end{equation}
and with an affine connection
${\hat \Gamma}^{\lambda}\,_{\mu\nu}$, satisfying the metricity
condition, and thereby related to the Christoffel
symbols and Cartan's torsion tensor as:
\begin{equation}
{\hat \Gamma}^{\lambda}\,_{\mu\nu}\,=\,\Gamma^{\lambda}\,_{\mu\nu}+
T^{\lambda}\,_{\mu\nu}.
\end{equation}
The torsion tensor is assumed to be
completely antisymmetric to fit the autoparallels
with the geodesics of the manifold.  The Hamiltonian for a dynamical
system is constructed as
\begin{equation}
H\,=\frac{1}{2}\,g^{\mu\nu}p_{\mu}p_{\nu}.
\label{gham}
\end{equation}
On the phase space, the expression for the covariant derivative with
torsion tensor is given by
\begin{equation}
{\hat D}_{\mu}F\,=\,\partial_{\mu}F
+p_{\lambda}{\hat \Gamma}^{\lambda}\,_{\mu\nu}
\frac{\partial F}{\partial p_{\nu}}
\end{equation}
where F is any function.  The Poisson-Dirac brackets
are expressed as
\begin{equation}
\{F,G\} \,=\,{\hat D}_{\mu}F \frac{\partial G}{\partial p_{\mu}}
-\frac{\partial F}{\partial p_{\mu}}{\hat D}_{\mu}G
-2\,p_{\lambda}\,T^{\lambda}\,_{\mu\nu}
\frac{\partial F}{\partial p_{\mu}}
\frac{\partial G}{\partial p_{\nu}}
\end{equation}
and the fundamental brackets are
\begin{equation}
\{q^{\mu},p_{\nu}\}\,=\,\delta^{\mu}_{\nu},
\qquad
\{p_{\mu},p_{\nu}\}\,=\,2\,p_{\lambda}\,T^{\lambda}\,_{\mu\nu}.
\end{equation}
In the phase space, the geodesic equations read 
\begin{equation}
{\dot q^{\alpha}}\,=\,\{q^{\alpha},H\}\,=\,p^{\alpha},
\qquad
{\dot p^{\alpha}}\,=\,\{p^{\alpha},H\}\,
=\,{\hat \Gamma}^{\lambda}\,_{\mu\nu}p^{\mu}p^{\nu}.
\end{equation}
The complete integrability of these equations are secured with the
existence of the Lax pair equations:~\cite{lax68}
\begin{equation}
{\dot L}=\{L,H\}=[L,A].
\label{dotl}
\end{equation}
Referring to~\cite{ros98}, it will be assumed that
$L$ is a homogeneous first order polynomial in momenta
$L^{\alpha}\,_{\beta}=L^{\alpha}\,_{\beta}\,^{\gamma}p_{\gamma}$.
The second matrix $A$ is also of the same form with respect to
the momenta
$A^{\alpha}\,_{\beta}=A^{\alpha}\,_{\beta}\,^{\gamma}p_{\gamma}$.
These third rank objects are referred as the Lax tensor and the Lax
connection, respectively.  After these preliminaries the brackets
$\{L,H\}$ are evaluated for the time derivative
of the Lax matrix 
\begin{equation}
{\dot L}^{\alpha}\,_{\beta}\,
=\,(L^{\alpha}\,_{\beta}\,^{(\mu}\,_{,\gamma}g^{\nu)\gamma}+
L^{\alpha}\,_{\beta}\,^{\gamma}{\hat \Gamma}^{(\mu}\,_{\gamma}\,^{\nu)})
p_{\mu}p_{\nu}.
\end{equation}
The right hand side of (\ref{dotl}) can be written as
\begin{equation}
(L^{\alpha}\,_{\gamma}\,^{\mu}\,A^{\gamma}\,_{\beta}\,^{\nu}\,-\,
A^{\alpha}\,_{\gamma}\,^{\mu}L^{\gamma}\,_{\beta}\,^{\nu})p_{\mu}p_{\nu}.
\end{equation}
Comparing the right hand side of (\ref{dotl}) with the left hand side,
it is now possible to make the following identification:
\begin{equation}
A^{\alpha}\,_{\beta}\,^{\gamma}\,
=\,{\hat \Gamma}^{\alpha}\,_{\beta}\,^{\gamma}\,
=\Gamma^{\alpha}\,_{\beta}\,^{\gamma}\,+\,T^{\alpha}\,_{\beta}\,^{\gamma}.
\end{equation}
In general, $B^{\alpha}\,_{\beta\gamma}$ in (\ref{one}) is an
arbitrary third rank tensor, unless some symmetry properties are 
imposed on it.  However, assuming it to be completely antisymmetric,
it can be identified with a geometrical object as
Cartan's torsion tensor.  Furthermore, with such an identification
the Lax tensor equations reduce to a compact form:
\begin{equation}
L_{\alpha\beta(\gamma;\delta)}\,=\,0
\label{master}
\end{equation}
where, the semicolon denotes the covariant differentiation with
respect to the affine connection.

Particularly, if $B_{\alpha\beta\gamma}\,=\,L_{\alpha\beta\gamma}$,
then the right hand side of (\ref{one}) vanishes.
In n-dimensional Euclidean or Minkowskian spacetimes they admit a
solution of the form:
\begin{equation}
L_{\alpha\beta\gamma}\,=\,F_{\alpha\beta\gamma\kappa}x^{\kappa}
+C_{\alpha\beta\gamma}.
\end{equation}
Here the tensor $F_{\alpha\beta\gamma\kappa}$ satisfies
$F_{\alpha\beta\gamma\kappa}=-F_{\beta\alpha\gamma\kappa}
=-F_{\alpha\beta\kappa\gamma}$, and  $C_{\alpha\beta\gamma}$
is a constant tensor.  If $B_{\alpha\beta\gamma}$ is completely
antisymmetric, then with
$C_{\alpha\beta\gamma}\,=\,\epsilon_{\alpha\beta\gamma}$, 
the equations (\ref{one}) and (\ref{master}) are equivalent.

\section{The Lax tensor equations on three dimensional spacetimes}
\noindent
Specifically, the Lax tensor can be split into completely
symmetric and antisymmetric parts
$L_{\alpha \beta \gamma}=S_{\alpha \beta \gamma}+R_{\alpha \beta\gamma}$.
For a more detailed analysis we are confined to three dimensions,
for the following reasons:~
First, the antisymmetric part of the Lax tensor becomes
proportional to $\epsilon_{\alpha \beta \gamma}$. Then equation
(\ref{master}) reduces to a simple form
\begin{equation}
S_{\alpha \beta \gamma ; \mu}=0
\label{symlax}
\end{equation}
for the symmetric part when $n < 4$, while for $n < 3$ 
the introduction of a completely antisymmetric third rank
tensor is not possible, to identify it with torsion.

In view of (\ref{symlax}), $S_{\alpha \beta \gamma}$ is a covariantly
constant tensor.  Spacetimes admitting such tensors are analyzed in the
context of recurrent tensors when
$T_{\alpha\beta}\,^{\gamma}=0$~\cite{will59}.
The integrability condition for (\ref{symlax}) can be expressed as:
\begin{equation}
S_{\mu[\alpha\beta}R^{\mu}\,_{\gamma]\,\rho\sigma}
  \,+\,
2\,\partial_{[\rho}(T_{\sigma][\alpha}\,^{\mu}S_{\beta\gamma]\mu})\,=\,0.
\label{sint}
\end{equation}
For simplicity in the notation and when no confusion is possible,
$[\,]$ either denotes antisymmetrization, or cyclic permutations.
After a detailed analysis of the above consistency condition for
an arbitrary diagonal metric with $T_{\alpha\beta}\,^{\gamma}=0$,
we have found that all six surviving components of the Riemann tensor
are equal and the surviving Lax tensor components are related as:
$S_{111}\,=\,S_{333}\,=\,2\,S_{123},\;S_{222}=-2S_{123}$.  However,
these restrictions turned out to be very severe on the manifold, yielding
the metric components to be constants.  Therefore, we can state that
the Lax equations are integrable if and only if the manifold is flat.

In the following we will present some examples for the
solutions of Lax equations in three dimensions.

\vspace*{0.325truein}
\subsection{Examples}
\noindent
\subsubsection{ The three dimensional Rindler system}
\noindent
One of the many possible generalizations of a two dimensional 
Rindler system to three dimensions can be obtained by assuming the
coordinates as
\begin{equation}
x=r \cosh \tau \cos \theta,\qquad y=r \cosh \tau \sin \theta,\qquad
t=r \sinh \tau
\end{equation}
with $0 \,< r\, <\, \infty, \quad 0 \,\leq \theta \, <\,2\pi, \quad
- \infty  \,< \tau \, <\, \infty  $.  The associated metric is
\begin{equation}
ds^{2}=dr^{2}+r^{2} \cosh^{2}\tau\,d\theta^{2}-r^{2}\,d\tau^{2}.
\end{equation}
The $\theta =0$ hypersurface defines the well-known two
dimensional Rindler system~\cite{hint97}.
The symmetric Lax tensor has ten components in three dimensions and 
without torsion the solutions of (\ref{symlax}) are found to be
\begin{equation}
\begin{array}{l}
S_{111}=3\exp(\tau)-3\exp(-\tau)+\exp(3\tau)-\exp(-3\tau), \\[2mm]
S_{113}=r\,(\exp(\tau)+\exp(-\tau)+\exp(3\tau)+\exp(-3\tau)),  \\[2mm]
S_{133}=r^{2}(-\exp(\tau)+\exp(-\tau)+\exp(3\tau)-\exp(-3\tau)), \\[2mm]
S_{333}=r^{3}(-3\exp(\tau)-3\exp(-\tau)+\exp(3\tau)+\exp(-3\tau)), \\[2mm]
S_{122}=\frac{1}{2}\,r^{2}\exp(-3\tau)(\exp(6\tau)
                  + \exp(4\tau) - \exp(2\tau) -1), \\[2mm]
S_{223}=\frac{1}{2}\,
   r^{3}\exp(-3\tau)(\exp(6\tau) + 3\exp(4\tau) + 3\exp(2\tau) + 1), \\[2mm]
S_{112}=0,\qquad S_{123}=0, \qquad S_{233}=0,\qquad S_{222}=0.
\end{array}
\end{equation}
There are several relations between the Lax tensors and second
rank Killing tensors.  A second rank Killing tensor satisfies
\begin{equation}
D_{[\mu}K_{\alpha\beta]}=0.
\label{kill}
\end{equation}
When the spacetime is flat a Lax tensor can be constructed
as:~\cite{bk99}
\begin{equation}
L_{\alpha\beta\gamma}=D_{\alpha}K_{\beta\gamma}-D_{\beta}K_{\alpha\gamma}.
\label{laxkill}
\end{equation}
A solution to (\ref{kill}), for this metric is found as
\begin{equation}
K_{11}=1, \qquad K_{22}=\tau,  \qquad
K_{33}=(1+r^{2})\,r^{2}\cosh^{2}\tau.
\label{kill1}
\end{equation}
When a second rank Killing tensor is non-degenerate, it can be considered
as a metric itself, defining a "dual" spacetime~\cite{holt96}.  
Although, three dimensional Rindler system defines a flat spacetime, 
its dual spacetime is curved, with a curvature scalar
$R= \frac{-2\,(8r^{2} + 9)}{(r^{2} + 1)^{2}}$.
In view of (\ref{laxkill}) and (\ref{kill1}) we obtain
\begin{equation}
L_{122}=3r^{3}\cosh^{2}\tau, \qquad L_{133}=-3r^{3}
\end{equation}
as the two surviving components of the Lax tensor.
Further relations between the Killing tensor and the Lax tensor
can be established as
$k_{\mu\nu}=S_{\mu\alpha\beta}S_{\nu}\,^{\alpha\beta}$.
Such a Killing tensor is trivial~\cite{book}.  Its surviving components
are:
\begin{equation}
\begin{array}{l}
k_{11}= 4\exp( - 2\tau)(\exp(4\tau) - 10\exp(2\tau) + 1),\\[2mm]
k_{13}= 4r\,\exp( - 2\tau)(\exp(2\tau) + 1)(\exp(\tau) + 1)
                (\exp(\tau) - 1),\\[2mm]
k_{22}= - 8r^{2}\,\exp( - 2\tau)(\exp(2\tau) + 1)^{2},\\[2mm]
k_{33}= 4r^{2}\,\exp( - 2\tau)(\exp(4\tau) + 10\exp(2\tau) + 1).
\end{array}
\end{equation}
The dual spacetime associated to this Killing tensor is flat.
\subsubsection{Ellipsoidal coordinates system}
\noindent
Separable coordinate systems in three-dimensional Minkowski
space label confocal surfaces of order two~\cite{hint97}.  In the
following coordinates will be denoted by $\mu,\nu,$ and $\rho$,
defined on the intervals
$-\infty\, <\,\nu \,< \,0\,<\, \rho \,< \,1\,<\,a\,<\, \mu \,< \infty  $,
where $a \in R$.
The metric corresponding of the Ellipsoidal coordinates systems,
falls into this class
\begin{equation}
g_{ij}=\frac{1}{4}diag\left[{(\mu-\nu)(\mu-\rho)\over\mu(\mu-1)(\mu-a)},
{(\nu-\mu)(\nu-\rho)\over\nu(\nu-1)(\nu-a)},
{(\rho-\mu)(\rho-\nu)\over\rho(\rho-1)(\rho-a)}\right].
\end{equation}
Non-degenerate Killing metrics corresponding
to the Ellipsoidal coordinate system are immediately
calculated as:
\begin{equation}
k_{ij}={1\over\mu\nu\rho}
diag\left[{(\mu-\nu)(\mu-\rho)\over(\mu-a)(\mu-1)},
{(-\mu+\nu)(\nu-\rho)\over(\rho-1)(\nu-a)},
{(-\mu+\rho)(-\nu+\rho)\over(\rho-1)(\rho-a)}\right].
\end{equation}
For $L_{\alpha\beta\gamma}=-L_{\beta\alpha\gamma}$
solutions to (\ref{master}) are found as:
\begin{equation}
\begin{array}{l}
L_{121}=(\mu-\rho)(\mu+\nu-2a)/\left[
       2\mu(\mu-1)(a-\rho)(\nu-a)(\mu+a)^2 \right],\\[2mm]
L_{122}=(\nu-\rho)(\mu+\nu-2a)/\left[
       2\nu(\nu-1)(a-\rho)(\nu-a)^2(a-\mu)\right],\\[2mm]
L_{131}=(\nu-\mu)(2a -\mu-\rho)/\left[
       2\mu(\mu-1)(a -\rho)(\nu-a)(a-\mu)^2\right],\\[2mm]
L_{133}=(\rho-\nu)(2a-\mu -\rho)/\left[
       2\rho(\rho-1)(a -\rho)^2(\nu-a)(a-\mu)\right],\\[2mm]
L_{232}=(\mu-\nu)(\nu +\rho-2a)/\left[
       2\nu(a-\rho)(\nu-a)^2(a-\mu)\right],\\[2mm]
L_{233}=(\mu-\rho) (\nu+\rho-2a)/\left[
        2\rho(\rho-1)(a -\rho)^2(\nu-a)(a-\mu)\right],\\[2mm]
L_{231}=0,\qquad L_{123}=0, \qquad L_{132}=0. \\[2mm]
\end{array}
\end{equation}

\subsubsection{The three-particle open Toda lattice}
\noindent
The dynamics of the three-particle open Toda Lattice
can be formulated through a purely kinetic Hamiltonian
\begin{equation}
H\,=\,\frac{1}{2}[(1+2a_{1}\,^{2})p_{1}\,^{2}+p_{2}\,^{2}
+(1+2a_{2}\,^{2})p_{3}\,^{2}].
\end{equation}
In view of (\ref{gham}) the components of the diagonal metric
are found to be
\begin{equation}
g_{11}=(1+2a_{1}\,^{2})^{-1}, \qquad g_{22}=1,\qquad
g_{33}=(1+2a_{2}\,^{2})^{-1}.
\label{tmet}
\end{equation}
At this point we refer to Sec. 2 and relax the totally
antisymmetric condition on the torsion tensor, but consider
it in its most general form, which is
$T_{\alpha\beta}\,^{\gamma}=-T_{\beta\alpha}\,^{\gamma}$,
still keeping the metricity condition.
Even with this relaxation the autoparalles are retained
on the manifold.  Now, the affine connection differs from
the Christoffel symbols by the contorsion tensor
$K_{\alpha\beta}\,^{\gamma}=-K_{\beta\alpha}\,^{\gamma}$ as:~\cite{hehl76}
\begin{equation}
{\hat \Gamma}_{\alpha\beta}\,^{\gamma}=\Gamma_{\alpha\beta}\,^{\gamma}
+K_{\alpha\beta}\,^{\gamma}
\end{equation}
whose relation to the torsion tensor is defined through
$K_{\alpha\beta}\,^{\gamma}:=
T_{\alpha\beta}\,^{\gamma}-T_{\beta}\,^{\gamma}\,_{\alpha}
+T^{\gamma}\,_{\alpha\beta}$.  The surviving components of
$L_{\alpha\beta}\,^{\gamma}(=L_{\beta\alpha}\,^{\gamma})$
are found
\begin{equation}
\begin{array}{lll}
L_{11}\,^{1}=g_{11}, & L_{12}\,^{1}=a_{1}\sqrt{g_{11}}, &
L_{22}\,^{2}=1,\\
L_{23}\,^{3}=a_{2}\sqrt{g_{33}}, &  L_{33}\,^{3}=g_{33}. &
\end{array}
\label{tlax}
\end{equation}
We give the surviving components of the contorsion tensor as:
\begin{equation}
K_{12}\,^{1}=a_{1}\sqrt{g_{11}}-2a_{1}^{2}g_{11}, \qquad
K_{23}\,^{3}=a_{2}\sqrt{g_{33}}-2a_{2}^{2}g_{33}.
\end{equation}
The tensorial Lax equation (\ref{one}) is satisfied when 
$L_{\alpha\beta}\,^{\gamma}$ is as in (\ref{tlax}) and
$B_{\alpha\beta}\,^{\gamma}=K_{\alpha\beta}\,^{\gamma}$
is as above.  Therefore, for this particular example
$B_{\alpha\beta}\,^{\gamma}$ can be interpreted as the
contorsion tensor.

\section{Conclusion}
\noindent
In this paper, we generalized the Lax tensor equations introduced by
Rosquist, by appropriately defining the Poisson brackets in the
presence of torsion.  This way, otherwise arbitrary tensors of these
equations can be identified with concrete geometrical objects, such as
the torsion or the contorsion tensor, when some relevant symmetry
properties are imposed on them.  The form of the equations are
considerably simplified, when $B^{\alpha}\,_{\beta\gamma}$ is 
completely antisymmetric.

We have also found the conditions when the Lax
equation on a three dimensional manifold admit solutions.    
We analyzed separable coordinates and the three-particle
open Toda lattice, in detail.

As was pointed in~\cite{book} Killing tensors can be trivial
or non-trivial.  A similar characterization arises when we 
investigate the solutions of the Lax tensor equations.  
If the Lax tensors satisfy
$g_{\mu\nu}=L_{\mu\alpha\beta}L_{\nu}\,^{\alpha\beta}$,
then they are non-trivial tensors.  

Further intriguing problems are to investigate the existence of the Lax
tensors, when the manifold admits Runge-Lenz symmetry, or to find
Lax tensors for superintegrable systems.  These problems are currently
under investigation~\cite{bb00}.
\noindent

\section*{Acknowledgments}
\noindent
One of us (D. B.) is grateful to Ashok Das for valuable discussions.
We would also like to thank to Y. G{\"u}ler for encouragements.


\begin{thebibliography}{000}
\bibitem{ros97} K. Rosquist, in {\em The Seventh Marcel Grossmann
Meeting.  On Recent Developments in Theoretical and Experimental General
Relativity, Gravitation and Relativistic Field Theories} {\bf 1}, 379.
eds. R. T. Jantzen and G. M. Keiser (World Scientific, Singapore, 1997)
\bibitem{ros98} K. Rosquist and M.Goliath {\em Gen. Rel. Grav.}
                {\bf 30}, 1521 (1998).
\bibitem{rosq3} M. Karlovini and K. Rosquist, preprint gr-qc/9807051.
\bibitem{patru} M. Goliath, M. Karlovini and K. Rosquist,
                               preprint solv-int/9810011.
\bibitem{cinci} K. Rosquist and M. Karlovini,
               {\em Journ. Math. Phys.} {\bf 41}, 370 (2000).
\bibitem{lax68} P. D. Lax, {\em Comm. Pure. Appl. Math.}
                           {\bf 21}, 467 (1968).
                A.M.Perelomov, {\em Integrable systems of classical
mechanics and Lie algebra} (I. Birkhauser, 1990)
\bibitem{holt96} R. H. Rietdijk and J. W. van Holten,
              {\em Nuc. Phys. B} {\bf 472}, 427 (1996).
\bibitem{bk99} D. Baleanu and A. (Kankanl{\i}) Karasu,
      {\em  Mod. Phys. Lett. A} {\bf 14}, 2587 (1999).
\bibitem{bochner} S. Bochner, {\em Ann. Math.} {\bf 49}, 379 (1948).
\bibitem{yano52} K. Yano, {\em  Ann. Math.} {\bf 55}, 328 (1952).
\bibitem{gib} G. Gibbons , R. H. Rietdijk and J. W. van Holten,
              {\em Nuc. Phys. B} {\bf 404}, 42 (1993).
\bibitem{visine} M. Visinescu and D. Vaman, {\em Phys. Rev. D}
                             {\bf 54 }, 1398  (1996).
\bibitem{carter} B. Carter, {\em Phys. Rev. D}
                        {\bf 16}, 3395 (1977).
\bibitem{holten95} J. van  Holten, {\em Phys. Lett. B}
                              {\bf 342} , 47 (1995).
\bibitem{das} S. Okubo and A. Das, {\em Phys. Lett. B}
                             {\bf 209}, 311 (1988).
\bibitem{will59} T. J. Willmore,
                  {\em An Introduction to Differential Geometry}
                 (Oxford University Press, Delhi, 1983).
\bibitem{book}  D. Kramer, H. Stepfani, E. Herlt and  M. Mac. Callum,
              {\em Exact Solutions of Einstein's Field Equations}
          (Cambridge University Press, Cambridge, 1980).
\bibitem{hint97} F. Hinterleitner, {\em Acta. Phys. Slovaca}
                      {\bf 47}, 157 (1997).  
\bibitem{hehl76} F. W. Hehl, P. von der Heyde and G. D. Kerlick
                {\em Rev. Mod. Phys.} {\bf 48}, 393 (1976).
\bibitem{bb00} D. Baleanu and S. Ba{\c s}kal, in preparation.

\end{thebibliography}
\end{document}